\documentclass[conference]{IEEEtran}
\IEEEoverridecommandlockouts
\usepackage{cite}
\usepackage{amsmath,amssymb,amsfonts}
\usepackage{algorithmic}
\usepackage{graphicx}
\usepackage{color}
\usepackage{textcomp}
\usepackage{tabularx}
\def\BibTeX{{\rm B\kern-.05em{\sc i\kern-.025em b}\kern-.08em
    T\kern-.1667em\lower.7ex\hbox{E}\kern-.125emX}}
\begin{document}

\title{Analytical Inverter Delay Modeling Using Matlab's Curve Fitting Toolbox}

\author{\IEEEauthorblockN{Walter Schneider}
\IEEEauthorblockA{\textit{Berufliche Oberschule Passau} \\
Passau, Germany\\
schneider@fos-bos-passau.de}
}

\maketitle

\begin{abstract}
This paper presents a new analytical propagation delay model for deep submicron CMOS inverters.
The model is inspired by the key observation that the inverter delay is a complicated function of several process parameters as well as load capacitance. These relationships are considered by fitting functions for each parameter derived from the {\it Curve Fitting Toolbox} in Matlab. Compared to SPICE simulations based on the BSIM4 transistor model, the analytical delay model shows very good accuracy with an average error less than $2\%$ over a wide range of process parameters and output loads. Hence, the proposed model can be efficiently used for different technology nodes as well as statistical gate delay characterisation.
\end{abstract}


\section{Introduction}
In CMOS digital design accurate timing characterisation of logic gates and digital circuits is essential.
For static timing analysis (STA) it is common practice that two dimensional lookup-tables are used for describing the delay of cells dependent from slope and load capacitance. Building such tables however is computationally quite expensive as lots of SPICE simulations are necessary. Therefore the development of accurate analytical timing models has been the subject of much research in the past.
As the CMOS inverter is an essential element in digital IC design analytical inverter delay models are of particular interest. Early models are based on the Shockley model for MOSFETs \cite{b1}, \cite{b14} which turn to be inaccurate in $90nm$ technologies and beyond. Later on Sakurai and Newton proposed an empirical compact MOSFET model known as Alpha-Power Law Model \cite{b3}, \cite{b4} which considers velocity saturation through parameter $\alpha$ (between $1$ and $2$). In the past several analytical delay formulas partly based on the Alpha-Power Low Model were introduced in \cite{b5}, \cite{b7}, \cite{b8} but there is still a lack in precise delay estimation. This is due to the complicated dependence of load capacitance, input slope, I/O coupling capacitance, short circuit current, velocity saturation, channel length modulation and drain-induced barrier lowering (DIBL) on delay.
In \cite{b6} the most promising approach for an analytical inverter delay model is presented which takes into account all mentioned physical effects. However for the drain current empirical parameters $K_{lin}$ and $K_{sat}$ are used. So the influence of process parameters like oxide thickness $t_{ox}$ or electron mobility $\mu$ on $K_{lin}$ and $K_{sat}$ is hidden. Furthermore process parameters tend to vary (see Table~\ref{tb:processParam}) within one technology node, so $K_{lin}$ and $K_{sat}$ cannot be treated as constants. This will lead to misleading results.

\renewcommand{\arraystretch}{1.3}
\begin{table}[h]
\label{tb:processParam}
\tabcolsep5mm
\begin{tabular}{|c|c|c|c|c|} \hline
Year & $L(nm)$ & $t_{ox}(nm)$ & $V_{th}(V)$  \\ \hline \hline
2001 & 180 &  4.5-5.5 & 0.39-0.43 \\ \hline
2003 & 130 &  3.5-4.0 & 0.35-0.40 \\ \hline
2004 & 90 &  1.6-3.0 & 0.25-0.40  \\ \hline
2007 & 65 &  1.5-2.0 & 0.20-0.35 \\ \hline
2009 & 45 &  1.0-1.4 & 0.20-0.35 \\ \hline
\end{tabular}
\centering
\caption{Evolution of nanometer CMOS characteristics \cite{b2}}
\end{table}

Our proposed method provides a compact analytical inverter delay model dependent from process parameters and load capacitance. In a first step we model the drain saturation current $I_{DnSat}$ (this current is sufficient for fast input ramps which will be the focus in this paper) as a function of process parameters. Then the effect of load capacitance on delay is investigated. By using the {\it Curve Fitting Toolbox} of Matlab \cite{b10} then we derive an analytical formula for the inverter delay dependent from $I_{DnSat}$ and load. 

The rest of the paper is organized as follows. In the next Section we review the technical background of inverter delay modeling. In Section III we demonstrate the procedure for obtaining an analytical saturation drain current followed by our analytical inverter delay model in Section IV. The experimental results are presented in Section V and finally we conclude in Section VI.

\section{Preliminaries}

\subsection{Theoretic Basics of Delay Modeling}

\begin{figure}[h]
\center
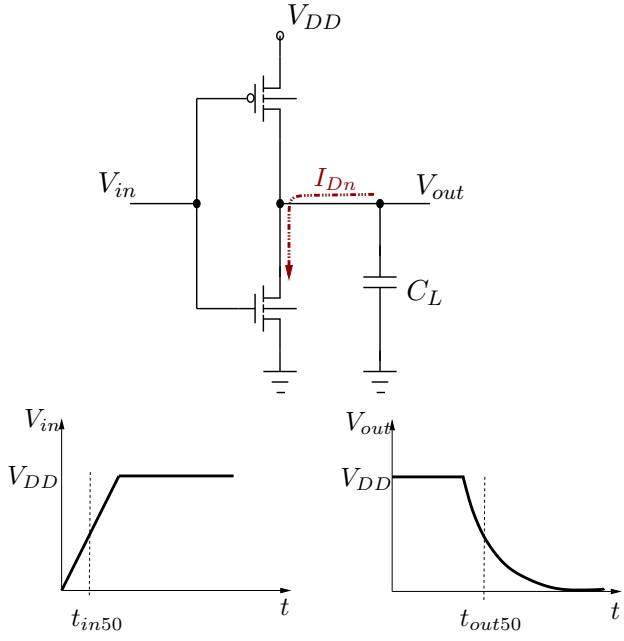
\caption{CMOS-inverter with load $C_L$ which is discharged through nMOS transistor for a rising input ramp}
\label{fig:inverter}       
\end{figure}

The average propagation delay $t_d$ of a CMOS inverter driving a load capacitance $C_{L}$ (Figure~\ref{fig:inverter}) is calculated by:
\begin{equation} \label{eq:averagedelay}
t_d = \frac{t_{pHL}+t_{pLH}}{2}
\end{equation}

where both components $t_{pHL}$ and $t_{pLH}$ are computed by $t_{out50} - t_{in50}$ \cite{b16}. $t_{out(in)50}$ refers to the time at which the output (input) voltage level reaches half of $V_{DD}$. 
When applying a rising input ramp the load capacitance is discharged by the drain current $I_{Dn}$ as shown in Figure~\ref{fig:inverter}.
The solution of the differential equation

\begin{equation}
C_L \cdot \frac{dV_{out}}{dt}=I_{Dn}
\end{equation}

yields the fall-delay $t_{pHL}$.

Under the assumption of a sufficient fast rising input ramp we can approximate $I_{Dn}$ with the drain saturation current $I_{DnSat}$. Then we have \cite{b8}
\begin{equation}
t_{pHL}=\frac{C_L \cdot V_{DD}}{2 \cdot I_{DnSat}}
\end{equation}

$I_{DnSat}$ is given according to \cite{b7}
\begin{equation}
I_{DnSat}=\frac{1}{2} \cdot \frac{\mu \epsilon_0 \epsilon_r}{t_{ox}} \cdot \frac{W}{L} \cdot \left(V_{GS}-V_{th} \right)^\alpha \cdot \left(1+\lambda V_{DS} \right)
\end{equation}

with $\mu$ as the electron mobility, $t_{ox}$ as the oxide thickness and $V_{th}$ as the threshold-voltage. $W$ is the channel-width, $L$ the channel-length, $\lambda$ the channel-length modulation factor and $\alpha$ the velocity saturation index. These parameters refer to the nMOS transistor. $V_{DS}, V_{GS}$ are the drain-source and gate-source voltage.

For deriving $t_{pLH}$ when applying a falling input ramp $C_L$ will be charged by $I_{Dp}$ of the pMOS transistor. Therefore the pMOS transistor parameters are needed. In the following we want to focus on $t_{pHL}$ but our method is also applicable for $t_{pLH}$.

\subsection{Determination of Drain Current: Mismatch between analytical model and SPICE simulation}
An accurate drain current model is crucial for an exact gate delay. A comparison between the analytical drain saturation current in eq.(4) and SPICE simulation is quite difficult due to the highly complex relationships between model parameters for SPICE and physical parameters used in eq.(4) \cite{b11}.

\begin{table}[h]
\label{tb:modelParam}
\tabcolsep1mm
\begin{tabular}{|c|c|c|c|} \hline
Parameter & for eq.(4) &  for SPICE & value  \\
 & (physical parameter) & (model parameter) &  \\ \hline \hline
mobility & $\mu$ & $U0$ & $550 \frac{cm^2}{Vs}$ \\ \hline
threshold-voltage & $V_{th}$ & $VTH0$ & $0.4V$ \\ \hline
oxide thickness & $t_{ox}$ & $TOXE$ & $3.0nm$  \\ \hline
channel-length & $L_{eff}$ & $L$ & $90nm$ \\ \hline
\end{tabular}
\centering
\caption{Model parameters for SPICE and process parameters}
\end{table}

\begin{figure}[h]
\centerline{\includegraphics[width=9.5cm]{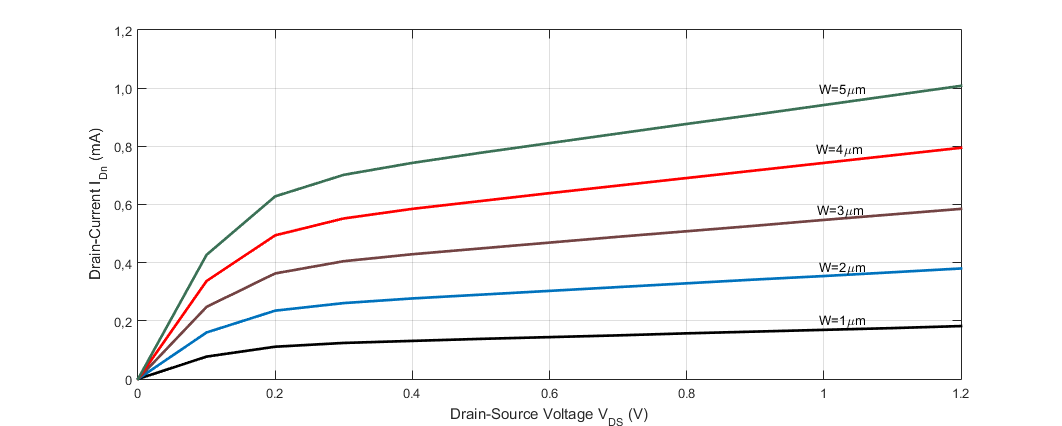}}
\caption{$I_{Dn}-V_{DS}$ characteristics of nMOS transistor dependent from transistor width $W$}
\label{fig:iddrain}
\end{figure}

Feeding the SPICE simulator with the model parameters as in Table~\ref{tb:modelParam} we get the following $I_{Dn}-V_{DS}$ characteristics of the nMOS transistor for different widths (see Figure~\ref{fig:iddrain}).

As for the gate delay computation with fast input ramps only the saturation drain current $I_{DnSat}$ is important we take $I_{DnSat}$ for $V_{GS}=V_{DS}=V_{DD}=1.2V$ from these curves and compare them to the analytical $I_{DnSat}$. From Figure~\ref{fig:compSPICE} we can conclude:

\begin{figure}[h]
\centerline{\includegraphics[width=9.5cm]{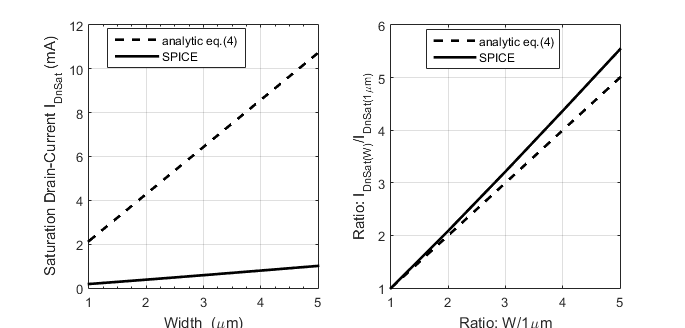}}
\caption{$I_{DnSat}$ with SPICE and eq.(4) (left), dependency of $I_{DnSat}$ on channel width $W$ (right)}
\label{fig:compSPICE}
\end{figure}

\begin{itemize}
\item The different nature of physical parameters (used in eq.(4)) and the model parameters for SPICE cannot be neglected. The model parameters, such as $U0$ or $VTH0$ must be converted to {\it effective} parameters like $\mu$ and $V_{th}$ in order to use the analytical model. Such complicated equations as described in the BSIM4 manual \cite{b11} with more than $200$ model parameters may lead to an impractical delay model.
\item The dependency of $I_{DnSat}$ from parameters is quite more difficult as stated in eq.(4). For example there is no real linear dependency from the channel width $W$ as seen in the right part of Figure~\ref{fig:compSPICE}. This is also for other parameters.
\end{itemize}

Many papers use the transconductance parameter $K$ empirical achieved by SPICE simulation (e.g. \cite{b6}) for their analytical models. Though a resonable accuracy is reached the dependency of $K$ from process/model parameters is lost.
To overcome this problem in the following Section we suggest a fitting-based approach to get an analytical expression for $I_{DnSat}$, which preserves the dependency on process/model parameters. \\

\section{Analytical Drain Saturation Current}

Starting with a specific technology node the following steps yield the desired analytical drain current equation. \\ 

\begin{figure}[h]
\centerline{\includegraphics[width=9.2cm]{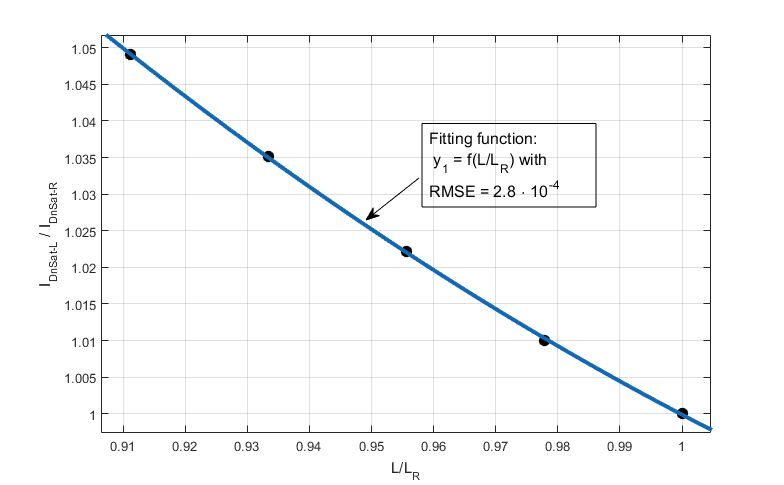}}
\caption{Quadratic fitting function $y_1$ which fits $\frac{I_{DnSat-L}}{I_{DnSat-R}}$ dependent from $\frac{L}{L_R}$.}
\label{fig:fit_l}
\end{figure}

\fbox{
\parbox{8cm}{
\begin{itemize}
\item[1.] Generate a {\it reference} saturation drain current $I_{DnSat-R}$ through SPICE simulation with reference values for the parameters $TOXE$, $VTH0$, $U0$, $L$ and $W$. 
\item[2.] Specify an interval for each parameter by using Table~\ref{tb:processParam} or the Predictive Technology Model PTM \cite{b9}. Subdivide each interval in equal steps, e.g. $L$ becomes a list $(L_1,L_2,...,L_n)$ consisting of $n$ different channel-lengths.
\item[3.] Generate a list of saturation drain currents for all values within the list for one parameter through SPICE simulation. All other parameters are set to their reference values. Refering to $L$ we get $n$ currents, denoted as $I_{DnSat-L}$.
\item[4.] Start the {\it Curve Fitting}-Toolbox from Matlab with the following inputs:
\begin{itemize}
\item {\it x-data} = ratio between varied parameter and reference parameter, e.g. $\frac{L}{L_R}$
\item {\it y-data} = ratio between simulated drain saturation current and $I_{DnSat-R}$, e.g. $\frac{I_{DnSat-L}}{I_{DnSat-R}}$
\end{itemize}
Choose an appropriate fitting function for the investigated parameter, e.g. a quadratic fitting function $y_1$ is used for $L$ which is illustrated in Figure~\ref{fig:fit_l}. 
\item[5.] Repeat the steps $3$ and $4$ for all other parameters. From the toolbox we get the fitting functions denoted as $y_2,y_3,y_4,y_5$.
\item[6.] The formula for the analytical drain saturation current is given as: 
\begin{equation}
I_{DnSat-A}=y_1 \cdot y_2 \cdot y_3 \cdot y_4 \cdot y_5 \cdot I_{DnSat-R}
\end{equation}
\end{itemize}
}
}

\vspace{0.5cm}
\section{Analytical Inverter Delay}
\begin{figure}[h]
\centerline{\includegraphics[width=9.2cm]{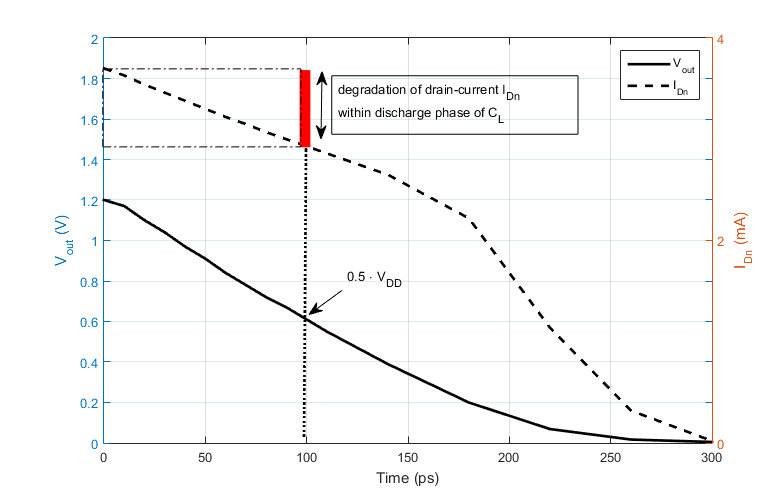}}
\caption{Dependency of the drain-current $I_{Dn}$ of the output voltage $V_{out}$ when discharging the load capacitance to $\frac{V_{DD}}{2}$}
\label{fig:iddischarge}
\end{figure}

Now for computing the delay $t_{pHL}$ as described in Section II we take the load capacitance $C_L$ into account. However using the simple eq.(3) is problematic since $I_{DnSat}$ is not really constant for the timing interval of discharging the load. This is illustrated in Figure~\ref{fig:iddischarge}.

\begin{figure}[h]
\centerline{\includegraphics[width=9.5cm]{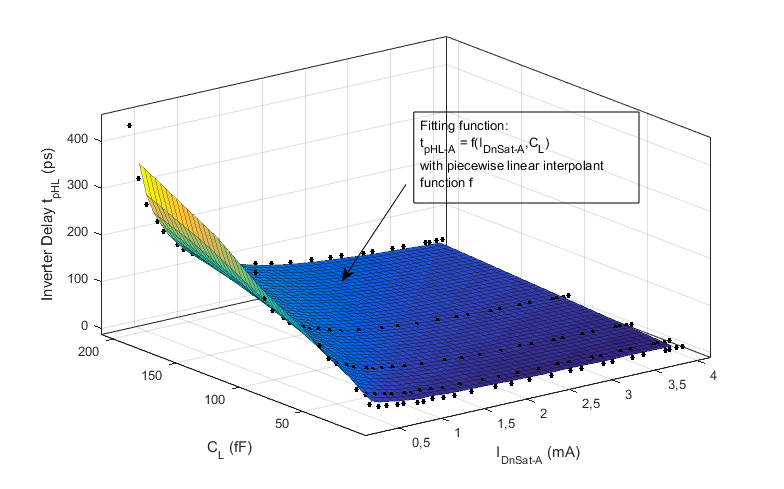}}
\caption{Piecewise linear interpolant fitting function $t_{pHL-A}$ which fits $t_{pHL}$ dependent from  $I_{DnSat-A}$ and $C_L$}
\label{fig:fitting2}
\end{figure}

In order to work with our analytical drain saturation current formula eq.(5), the following procedure is suggested.
\begin{itemize}
\item Simulate the delay $t_{pHL}$ with SPICE for the parameter constellations as in Section III and different loads $C_L$.
\item Compute $I_{DnSat-A}$ using eq.(5) with these parameter constellations.
\item Start the {\it Curve Fitting}-Toolbox from Matlab with the following inputs:
\begin{itemize}
\item {\it x-data} = $I_{DnSat-A}$
\item {\it y-data} = loads $C_L$
\item {\it z-data} = $t_{pHL}$ values after SPICE simulation
\end{itemize} 
This means $t_{pHL}$ is a two-dimensional plane dependent from drain current and load capacitance as illustrated in Figure~\ref{fig:fitting2}. By choosing an appropriate fitting function we get the desired analytical model denoted as $t_{pHL-A}$.
\end{itemize}

{\it Note:} A visual examination of a fitted curve/plane which is displayed in the {\it Curve Fitting App} of Matlab should always be the first step in the evaluation of the fitting quality. Beyond the toolbox generates a {\it Goodness-of-Fit} Output, which provides goodness-of-fit statistics like the sum of squares due to error (SSE) or the root mean square error (RMSE) \cite{b10}, \cite{b15}.

\begin{table*}[t]
\tabcolsep0.1mm
\begin{tabular}{|c||c|c|c|c|c||c||c||c|c||} \hline
Test No. & \multicolumn{5}{c||} {Process/Model Parameters} &  {Simulated Delay (ps)} &  {Analytical Delay (ps)} & {Maximum Error} & {Average Error} \\
 &  $L (nm)$  & $W (\mu m)$ &  $TOXE (nm)$ & $VTH0 (V)$ & $U0 (\frac{cm^2}{Vs})$  &  &  & (\%) & (\%) \\ \hline \hline
1 & $80.8$ &  $2.8$  & $1.78$ & $0.27$ & $538$ & $(13.4,21.9,38.1,51.8,73.4)$ & $(14.1,22.5,38.1,51.9,72.4)$ & $5.0$ & $1.8$ \\ \hline
2 & $80.8$ & $2.8$ & $2.40$ & $0.37$ & $540$ & $(18.6,32.0,57.4,80.0,114.4)$ & $(17.9,31.1,55.9,77.5,111.3)$ & $3.9$ & $3.1$ \\ \hline
3 & $80.8$ & $1.34$ & $1.78$ & $0.27$ & $542$ & $(19.5,37.4,71.2,100.6,146.4)$ & $(20.0,37.8,71.3,100.1,145.4)$ & $2.5$ & $1.0$ \\ \hline
4 & $80.8$ & $4.65$ & $1.78$ & $0.27$ & $538$ & $(11.1,16.1,25.8,34.3,47.0)$ & $(11.1,16.1,25.7,33.8,46.6)$ & $1.5$ & $0.6$ \\ \hline
5 & $85.3$ & $2.8$ & $1.78$ & $0.33$ & $542$ & $(15.2,24.1,42.3,58.0,82.2)$ & $(15.1,25.4,43.1,59.1,82.8)$ & $5.1$ & $2.0$ \\ \hline
6 & $85.3$ & $1.34$ & $1.62$ & $0.37$ & $542$ & $(22.7,43.0,80.8,114.0,165.5)$ & $(22.8,44.0,83.9,117.5,170.5)$ & $3.7$ & $2.5$ \\ \hline
7 & $87.8$ & $2.8$ & $1.78$ & $0.37$ & $540$ & $(15.9,26.3,46.2,63.3,89.9)$ & $(15.6,26.3,46.1,63.4,89.9)$ & $1.9$ & $0.4$ \\ \hline
8 & $87.8$ & $1.34$ & $1.78$ & $0.27$ & $538$ & $(20.2,38.6,73.2,103.5,150.5)$ & $(21.1,39.7,74.9,107.0,153.0)$ & $4.3$ & $2.8$ \\ \hline
9 & $87.8$ & $4.65$ & $2.26$ & $0.37$ & $542$ & $(14.4,22.5,37.0,49.9,69.6)$ & $(13.8,21.2,35.9,48.5,68.2)$ & $6.1$ & $4.0$ \\ \hline
10 & $89.3$ & $2.8$ & $2.07$ & $0.30$ & $538$ & $(15.5,26.1,45.6,62.7,89.4)$ & $(15.6,26.2,45.9,63.0,89.6)$ & $0.7$ & $0.5$ \\ \hline
\end{tabular}
\centering
\caption{Comparison between SPICE inverter delay and Analytical delay model for different process/model parameters and load capacitances}
\label{tb:invDelay}
\end{table*}

\section{Experimental results}

\subsection{Validation of drain current model}
\begin{figure}[h]
\centerline{\includegraphics[width=9.5cm]{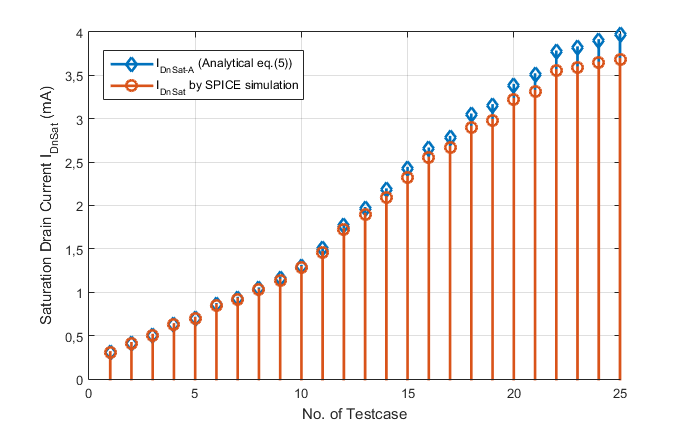}}
\caption{Comparison between simulated and analytical drain saturation currents}
\label{fig:drainCurrent_r}
\end{figure}

We start with specifying a list of values (valid for a $90nm$ technology node) for each parameter as described in Section III: $L=(90,88,86,84,82)nm$, $W=(1,2,3,4,5)\mu m$, $TOXE=(3,2.5,2,1.6)nm$, $VTH0=(0.4,0.35,0.3,0.25)V$ and $U0=(550,540,530,520)\frac{cm^2}{Vs}$.
Then $22$ SPICE simulations are needed to get the drain saturation currents. Together with one additional simulation for the reference drain current $I_{DnSat-R}$ (using the reference values for the parameters according to Table~\ref{tb:modelParam}) Matlab's {\it Curve Fitting} Toolbox can be feeded. It turns out that for each parameter a second order polynomial is an appropriate fit with a RMSE below $0.01\%$ in any case. Computing $I_{DnSat-A}$ with eq.(5) with all possible combinations of the parameter values listed above leads to $1600$ different values. Randomly we choose a subset of $25$ testcases. Figure~\ref{fig:drainCurrent_r} illustrates the quite good accordance of simulated and calculated saturation drain currents. The maximum error is smaller than $7 \%$, the average error is $3.0 \%$.

\subsection{Validation of inverter delay model}
First we take the parameter settings for the mentioned $25$ testcases and obtain $5$ inverter delays $t_{pHL}$ due to $C_L=(10,20,50,100,200)fF$ for each testcase by SPICE. Together with the values of $I_{DnSat-A}$ for these testcases, Matlab's {\it Curve Fitting} Toolbox generates a piecewise linear interpolant fit $t_{pHL-A}$ with $125$ coefficients.
Then we randomly generate $5$ values for each parameter with the Matlab routine {\it rand}, e.g.: $l_{var}=(90nm-10nm) \cdot rand(5,1)+80nm$. All values are within the lists as specified in Section V.A.  
Also we use $5$ different load capacitances in the interval $[10,200]fF$, namely $(13,37,83,123,185)fF$.

For each parameter combination and load capacitance we compute the analytical inverter delay using the fitting function $t_{pHL-A}$. Overall we get $5^6=15625$ values. In order to avoid simulating each case with SPICE, we choose $10$ different parameter combinations. For each combination SPICE reports the delay for all five load capacitances.
In Table~\ref{tb:invDelay} it is illustrated that the maximum error between SPICE and analytical model is about $6\%$. The average error for all cases is less than $2\%$.

\subsection{Comparison to other analytical models}
\begin{figure}[h]
\centerline{\includegraphics[width=10cm]{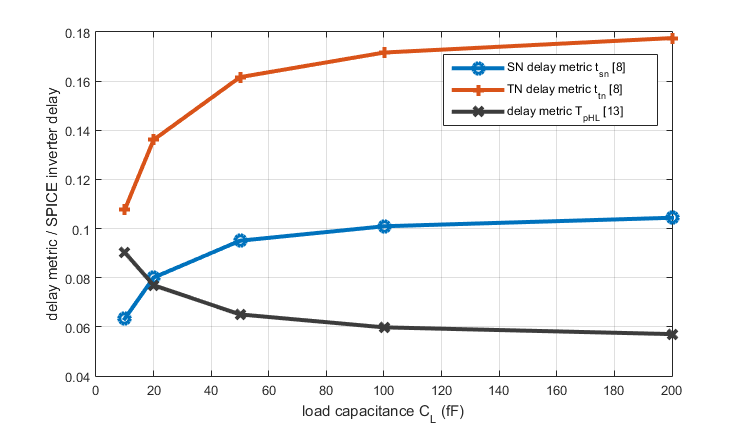}}
\caption{Comparison of delay metrics from \cite{b7},\cite{b12} and SPICE simulation}
\label{fig:delayMetric}
\end{figure}

Finally we want to evaluate the quality of recently published inverter delay equations. As reference we use the simulated SPICE delays for testcase No.1 and the loads $C_L$ as above. Both the Sakurai-Newton (SN) delay metric and the Taur-Ning (TN) delay metric from \cite{b7} underestimate the delay compared to SPICE. This underestimation becomes smaller with increasing loads. The delay $T_{pHL}$ from eq.(6) in \cite{b12} also underestimates the inverter delay, however the underestimation becomes smaller with decreasing loads. 
This is illustrated in Figure~\ref{fig:delayMetric}.
One can argue that in all cases this enormous discrepancy is caused by the mismatch of model and process parameters. But on the other hand a simple additional fitting factor would not solve the problem due to its dependency on $C_L$.

\section{conclusion}
In this paper we propose a novel analytical CMOS inverter delay model. The main advantage of this approach compared to previous related work is the more detailled study of the drain current especially the transconductance parameter dependent from process parameters. The model generation is both simple because of curve fitting and - apart from some necessary SPICE simulations - computationally inexpensive. Furthermore our method is applicable to any technology node. Experimental results for a wide range of different parameter values and load capacitances have shown that the worst case error is smaller than $7\%$. It is expected that our model is suitable to consider process variations in future technologies.


\begin{thebibliography}{00}
\bibitem{b1} Y. Taur and T. H. Ning, {\it Fundamentals of Modern VLSI Devices}, Cambridge University Press 1998.
\bibitem{b14} J. M. Rabaey, A. Chandrakasar and B. Nikolic, {\it Digital Integrated Circuits}, Prentice-Hall, 2003.
\bibitem{b2} E. Maricau and G. Gielen, {\it Analog IC Reliability in Nanometer CMOS}, Analog Circuits and Signal Processing, Springer Science+Business Media New York, 2013.
\bibitem{b3} T. Sakurai and A. R. Newton, {\it Alpha-power law MOSFET model and its applications to 'CMOS inverter delay and other formulas}, IEEE Journal of Solid State Circuits, vol. 25, no. 2, pp. 584-594, April 1990.
\bibitem{b4} T. Sakurai and A. R. Newton, {\it A simple MOSFET model for circuit analysis}, IEEE Transactions on Electron Devices, vol. 38, no. 4, pp. 887-894, April 1991.
\bibitem{b5} Y. Wang and M. Zwolinski, {\it Analytical transient response and propagation delay model for nanoscale CMOS inverter}, Proceedings of International Symposium on Circuits and Systems (ISCAS), pp. 2998-3001, 2009.
\bibitem{b6} F. S. Marranghello, A. I. Reis and R. P. Ribas, {\it Improving Analytical Delay Modeling for CMOS Inverters}, Journal of Integrated Circuits and Systems, vol. 10, no. 2, pp. 123-134, 2015.
\bibitem{b7} A. Ramalingam, S. V. Kodakara, A. Devgan and D. Z. Pan, {\it Robust Analytical Gate Delay Modeling for Low Voltage Circuits}, Proceedings of the 2006 confernce on Asia South Pacific design automation, pp. 61-66, Yokohama, Japan, 2006.
\bibitem{b8} N. Chandra, A. K. Yati and A.B. Bhattacharyya, {\it Extended-Sakurai-Newton MOSFET Model for Ultra-Deep-Submicrometer CMOS Digital Design}, 22nd International Conference on VLSI Design, 2009.
\bibitem{b9} W. Zhao and Y. Cao, {\it New generation of predictive technology model for sub-45nm early design exploration}, IEEE Transaction on Electron Devices, vol. 53, no. 11, pp. 2816-2823, 2006. Available online at: http://ptm.asu.edu.
\bibitem{b10} @ 1994-2017 The MathWorks, Inc., {\it Curve Fitting Toolbox}, online available at: https://de.mathworks.com/help/curvefit/index.html, 2017.
\bibitem{b11} C. Hu, S. Salahuddin, Y. Singh, S. S. Parihar and C. K. Dabhi, {\it BSIM 4.8.1 MOSFET Model, User's Manual}, Department of Electrical Engineering and Computer Sciences, Berkeley 2017.
\bibitem{b12} S. Gummalla, A. R. Subramaniam, Y. Cao and C. Chakrabarti, {\it An Analytical Approach to Efficient Circuit Variability Analysis in Scaled CMOS Design}, 13th International Symposium on Quality Electronic Design, 2010.
\bibitem{b15} P. Huber and E. Ronchetti, {\it Robust statistics}, Wiley New York, 2009.
\bibitem{b16} N. Weste and D. Harris, {\it CMOS VLSI Design 4th edition}, Boston: Addison-Wesley, 2010.

\end{thebibliography}
\end{document}